\documentclass[10pt,twocolumn,brazil,english,aps,pre]{revtex4-1}
\usepackage{mathptmx}

\usepackage[T1]{fontenc}
\usepackage[latin9]{inputenc}
\usepackage{color}
\usepackage{epsfig}
\usepackage{prettyref}
\usepackage{textcomp}
\usepackage{amssymb}
\usepackage{amsmath}



\makeatother

\usepackage{babel}

\begin{document}


\title{Energy Dissipation Via Coupling With a Finite Chaotic
Environment}

\author{M. A. Marchiori and M. A. M. de Aguiar}

\affiliation{Instituto de F\'isica ``\emph{Gleb Wataghin}", Universidade
Estadual de Campinas (UNICAMP),13083-859, Campinas, Brazil}

\begin{abstract}
We study the flow of energy between a harmonic oscillator (HO) and an
external environment consisting of $N$ two-degrees of freedom
non-linear oscillators, ranging from integrable to chaotic according to
a control parameter. The coupling between the HO and the environment is
bilinear in the coordinates and scales with system size as
$1/\sqrt{N}$. We study the conditions for energy dissipation and
thermalization as a function of $N$ and of the dynamical regime of the
non-linear oscillators. The study is classical and based on single
realization of the dynamics, as opposed to ensemble averages over many
realizations. We find that dissipation occurs in the chaotic regime for
a fairly small $N$, leading to the thermalization of the HO and
environment a Boltzmann distribution of energies for a well defined
temperature. We develop a simple analytical treatment, based on the
linear response theory, that justifies the coupling scaling and
reproduces the numerical simulations when the environment is in the
chaotic regime.
\end{abstract}

\maketitle

\section{Introduction}

Dissipation is usually modeled by coupling a {\it central system} to an
external {\it environment}. The environment can be represented by an
infinite collection of modes (either harmonic or spin-like) at a given
temperature, or by few chaotic degrees of freedom. Models in the first
category include the Caldeira--Leggett (CL) \cite{caldeira}, where the
environment is represented by $N$ harmonic oscillators with a linear
distribution of frequencies. In the thermodynamic limit
$N\rightarrow\infty$, the central system obeys the classical Langevin
equation, exhibiting exponential dissipation subjected to fluctuating
forces.

In the case of small environments, the exchange of energy with the
central system exhibits large fluctuations and equilibration is achieved
only after averaging over an ensemble of realizations, usually of the
micro-canonical type. Wilkinson \cite{wilkinson}, and later Berry and
Robbins \cite{berry}, studied the coupling of a fast chaotic environment
to a slow central system using the properties of adiabatic invariants.
They showed that the adiabatic invariance of the chaotic energy shell
\cite{ott,brown,brown1} leads, in first order, to a Born-Oppenheimer
conservative type of reaction on the central system. Second order
corrections lead to a force proportional to the velocity whose symmetric
part corresponds to friction and whose antisymmetric part acts as a
geometric magnetism  \cite{wilkinson,berry}. Since the publication of
these pioneering works, a large number of papers has explored the
possibility of dissipation caused by small environments, both classically
\cite{jarzynski,tulio,bonanca1,jane1,jane2} and quantum mechanically
\cite{cohen1,cohen2,cohen3,cohen4,auslaender,jalabert,karkuszewski,benenti,bonanca2,chou,pereverzev},
where decoherence is the main focus.

Although the approach to dissipation via ensemble average over small
environments has lead to several interesting results, it cannot account
for situations where a single experiment is performed and no averages
take place. This is, for example, the case of the $C_{60}$ molecule
interacting with its own internal degrees of freedom. As pointed out in
\cite{arndt}, couplings of these molecules to the environment or to
their own internal degrees of freedom play an essential role in the
appearance of decoherence, destroying self-interferences in a double
slit experiment, and deserve a careful analysis. When the environment
has relatively few degrees of freedom it is not clear whether they lead
to standard dissipation effects over the collective degrees of freedom.

The main goal of this paper is to study dissipation from the classical
point of view as induced by small environments without resorting to
ensemble averages. To distinguish this process from those obtained
after averaging, we term it {\it effective dissipation}. We follow the
general approach of Feynman and Vernon \cite{feynman} where the system
of interest, or {\it central system}, has much fewer degrees of freedom
than the environment, which, however, is also small. In our simulations
the central system is a harmonic oscillator with a single degree of
freedom whereas the environment has $2N$ freedoms. We study the system
as a function of $N$ and of its dynamical character, regular or
chaotic. We wish to determine the minimum value of $N$ for which the
environment behaves as a thermal bath, inducing irreversible
dissipation on the harmonic oscillator.

\section{Model}

We consider a globally conservative system governed by the Hamiltonian
\begin{equation}
H=H_{HO}+H_{E}+\lambda_{N}V_{I}
\label{eq1}
\end{equation}
where
\begin{eqnarray}
H_{HO} & = & \frac{p^{2}}{2m}+\frac{m\omega_{o}^{2}q^{2}}{2},
\label{eq:oscillator}\\
H_{E} & = & \sum_{n=1}^{N}\left[\frac{p_{x_{n}}^{2}+p_{y_{n}}^{2}}{2}+
\frac{a}{4}(x_{n}^{4}+y_{n}^{4})+\frac{x_{n}^{2}y_{n}^{2}}{2}\right]
\equiv \sum_{n=1}^{N} H_{Cn},
\label{eq:quartic system}\\
V_{I} & = & \sum_{n=1}^{N}qx_{n}.
\label{eq:coupling}
\end{eqnarray}

The environment $H_E$ is modeled by the finite sum of identical Quartic
Systems (QS) \cite{carnegie}. Each QS has two degrees of freedom and is
invariant under a scaling transformation in such a way that the
dynamics in all energy shells are the same. This property allows us to
adjust the QS's energy and time scales without changing its dynamical
regime, which is solely controlled by the parameter $a$. The QS is
integrable for $=1.0$, strongly chaotic for $a \lesssim 0.1$ and mixed
for intermediate values.

In addition to the dynamical regime, the number of QS's coupled to the
harmonic oscillator (HO) plays a crucial role in the dynamics of
dissipation. In order to compare results for different values of $N$,
the coupling parameter $\lambda_{N}$, as seen in equation \eqref{eq1},
has to be properly normalized with the size of the environment, so that
the effective coupling becomes independent of $N$ for large $N$. We use
$\lambda_{N}=\lambda/\sqrt{N}$ and derive this scaling later using
linear response theory.

We want to study the behavior of the system for a single global initial
condition and we consider two types of initialization. In both cases
the HO starts from $q=0$ and $p=\sqrt{2mE_o}$ with $m=1$,
$\omega_{o}=0.3$ and $E_{o}=10.0$. In the first type, the initial
condition for each QS is chosen randomly at a fixed energy shell $E_C$.
This choice is related to the microcanonical distribution
\begin{equation}
\rho=\frac{1}{Z}\prod_{n=1}^{N}\delta(H_{Cn}-E_{Cn}),
\label{eqdist}
\end{equation}
where $E_{Cn}=E_C$, for $n=1,\ldots,\, N$. We refer to it as {\it
pseudo-microcanonical}, to distinguish it from the real microcanonical
distribution which is uniform over the entire energy shell of the $N$
QS's.

The second type of initialization corresponds to pick a different
energy shell $E_{Cn}$ for each QS from the Boltzmann distribution $p(E)
=\exp{(-E/\bar{E}_{QS})}/\bar{E}_{QS}$ and we call it {\it pseudo-canonical}. The
initial condition for each QS is again chosen randomly at the
corresponding energy $E_{Cn}$.

\section{Numerical results}

In all simulations the coupling parameter will be fixed at
$\lambda=0.01$. We consider initially the pseudo-microcanonical
distribution with $E_C=0.01$.

\begin{figure}[ht]
\begin{centering}
\includegraphics[scale=0.35]{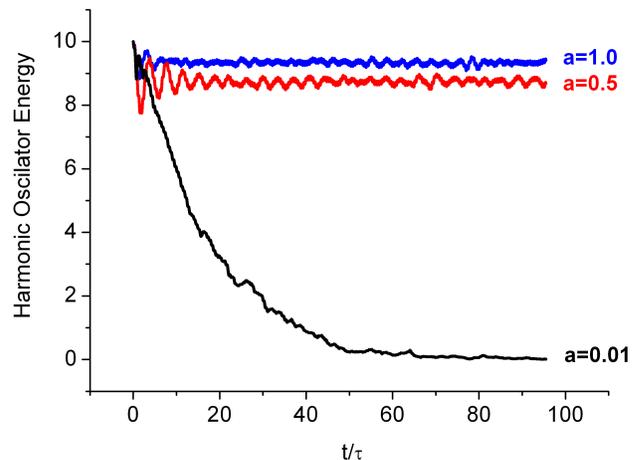}
\par
\end{centering}
\caption{(Color online) Energy of the HO when coupled to $N=100$ QS in
distinct dynamical regimes: integrable, $a=1.0$ (blue); mixed, $a=0.5$
(red); chaotic, $a=0.01$ (black).}
\label{fig1}
\end{figure}

We first study the role of chaos in promoting dissipation. Fixing
$N=100$ we integrate Hamilton's equations for different values of the
parameter $a$ and compute the energy of the harmonic oscillator $H_O$
as a function of the time. Time is measured in units of the HO's
period, $\tau=2\pi/\omega_{o} \approx 20.94$. Fig.\ref{fig1} shows
the results of simulations for a single numerical realization for three
values of $a$.

In the integrable regime, $a=1.0$, the effect of the environment is
minimal, causing a small decrease in the oscillator's energy and
inducing small amplitude, noisy, oscillations. In the mixed case,
$a=0.5$, the energy loss is slightly larger and the oscillations are
more clearly visible, particularly at shorter times. At long times the
energy fluctuates around $0.8E_{o}$. We refer to the papers
\cite{jane1,jane2} for numerical results on analogous systems, computed
via ensemble averages, in the regular and in mixed regimes. The authors
describe in great detail the dependence of the dissipation rate on the
number of degrees of freedom of the environment in these regimes.

In the chaotic regime the energy of the HO shows a clear exponential
decay for $N=100$ QS's. Similar results were obtained in
\cite{bonanca1} with $N=1$ but performing ensemble averages over
several thousands of trajectories. The results displayed in Fig.\ref{fig1} are for a single trajectory. Changing the initial conditions
in the environment, but keeping the same microcanonical constraint,
produces energy curves that differ only in the details of the
fluctuations, but not in their main qualitative features.

\begin{figure}[ht]
\begin{centering}
\includegraphics[scale=0.35]{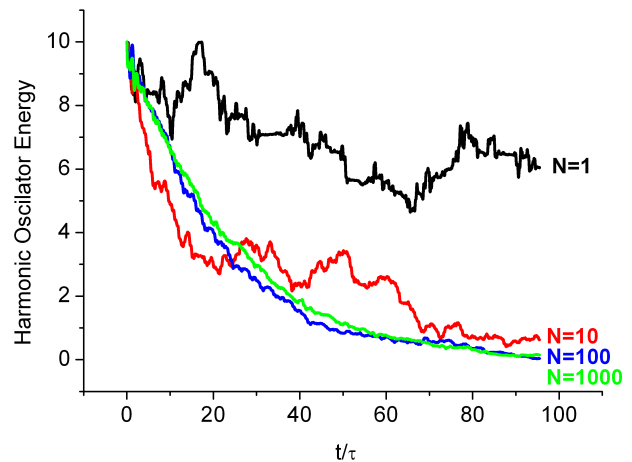}
\par
\end{centering}
\caption{Dissipation curve as function of the number
of QS in the environment.}
\label{fig2}
\end{figure}

The number of degrees of freedom of the environment is of fundamental
importance to the effective dissipative behavior. The exponential decay
observed in Fig.\ref{fig1} occurs only if $N$ is sufficiently large.
Fig.\ref{fig2} shows the typical behavior of $H_O(t)$ for different
values of $N$.

For small environments QS $(1\leq N\leq6)$, the fluctuations are so
large that dissipation cannot be characterized for each numerical
realization. In these cases, and in particular for $N=1$, exponential
dissipation can be observed only in an ensemble averaged treatment of
the energy \cite{bonanca1}.

Increasing the number of QS's to $7\leq N\leq20$ makes the decreasing
behavior of the HO energy more evident, although the fluctuations are
still large. Finally, for the parameters of this simulation, the curves
converge to a $N$-independent exponential decay $H_O(t) \simeq
e^{-\gamma t}$ for $N > N_0 \simeq 100$. Such convergence occurs only
if the coupling constant scales properly with $N$. Clearly
$\lambda_{N}$ has to decrease with the number of QS in the environment,
otherwise the interaction Hamiltonian $H_{I}$ would diverge as
$N\rightarrow\infty$. Only if $\lambda_{N}=\lambda/\sqrt{N}$, the
effect of the environment over the HO converges as $N$ becomes large,
as shown in Fig.\ref{fig2}. This scaling of the coupling constant will
be demonstrated using linear response theory.

Environments with $N \ge N_0$ behave as infinite reservoirs. Clearly,
$N_0$ depends on the model parameters and, in particular, on $a$, that
controls the degree of chaos in the QS. Indeed, dissipation is intimately
associated with correlation functions that contain information about the
memory of the environment. Increasing the phase space region covered by
chaos (by decreasing $a$) makes these correlation functions decay
faster, leading to a faster transfer of energy from the oscillator to the
environment, even for small values of $N$. This makes $N_0$ smaller for
smaller values of $a$.

The flow of energy between the HO and environment also depends on the
relative time scales of the HO and QS. Increasing the HO frequency
($\omega_{o}$) decreases the relative velocity of the two subsystems,
leading to a less efficient energy dissipation. However, since the QS is
scalable, we can always change its energy in such a way to readjust its
relative velocity.

\begin{figure}[ht]
\begin{centering}
\includegraphics[scale=0.35]{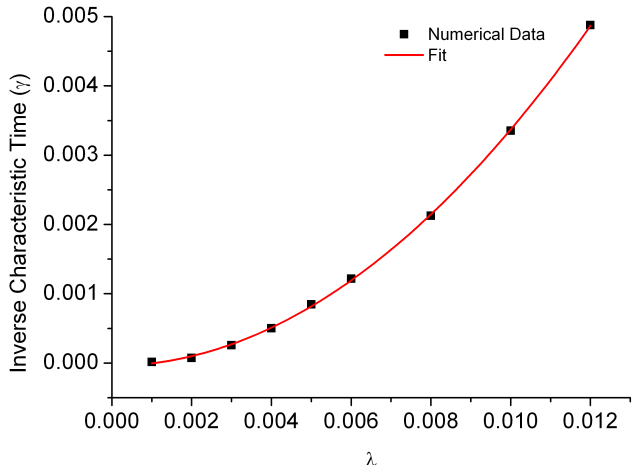}
\par
\end{centering}
\caption{Inverse characteristic time ($\gamma)$ as a function of the
coupling constant ($\lambda)$ for $N=100$ and $a=0.01$. The curve is
well fitted by a quadratic function.}
\label{fig3}
\end{figure}

The decay rate $\gamma$, or the characteristic time $t_{diss} = 1/\gamma$,
also changes with $\lambda$, as shown in Fig.\ref{fig3} for $N=100$
and $a=0.01$. The dependence is well fitted by a quadratic law, a
result that will also be demonstrated using linear response theory.

The numerical results in Fig.\ref{fig2} show that small chaotic
environments may behave as infinite reservoirs, absorbing energy from
the HO. This behavior can be further characterized by associating a
temperature to the full system after its equilibration. In order to do
so, we look at the energy distribution of the HO and the environment at
equilibrium. This can be done in two different ways: the first is to
count, at a fixed time, the number of QS's with energy in the interval
$(E,E+dE)$ for a large number of initial conditions (ensemble average).
The second possibility is to count, for a single realization, the
number os QS's with energy in the interval $(E,E+dE)$, starting at some
time after equilibration and repeating the procedure several times at a
pre-specified time step (time average). The same procedures can be
applied to the energy of the HO. The statistical distribution of these
results will be the same if the system is ergodic. We performed
numerical calculations using both the time and ensemble averages and
found very similar results, indicating that both the environment and
the HO are indeed ergodic in equilibrium.

\begin{figure}[ht]
\begin{centering}
\includegraphics[scale=0.35]{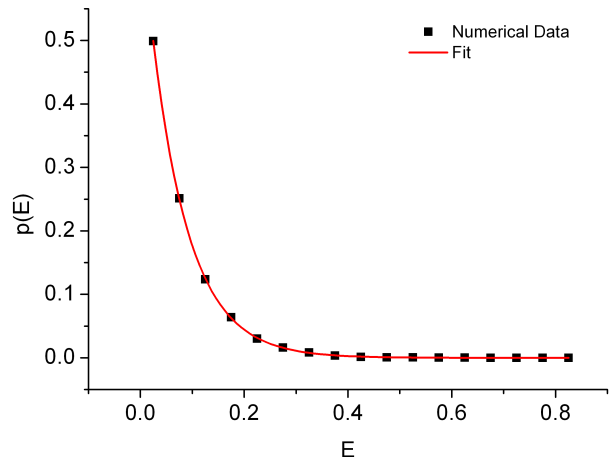}
\par
\end{centering}
\caption{Energy distribution for HO with $N=100$ QS's. The
temperature is $\bar{E}_{HO}=0.0723\pm0.0002$. Energy was sampled $100,000$
times at equal intervals.}
\label{fig4}
\end{figure}

In order to define a temperature for the environment we use the general
equipartition theorem \cite{tolman} to show that $\bar{E}_{QS}=
\frac{3}{2}\bar{E}_{HO}$. The energy distribution of the environment
for $N=100$ is well fitted by $\exp ({-E/\bar{E}_{QS}})$ with
$\bar{E}_{QS}\approx 0.1094$. The value of $\bar{E}_{QS}$ is a measure
of the environment's temperature. Fig.\ref{fig4} shows that the
energy distribution of the HO also follows a Boltzmann-like
distribution,
\begin{equation}
p_{OH}(E)= \frac{1}{\bar{E}_{HO}} e^{-E/\bar{E}_{HO}}
\label{boltz}
\end{equation}
where $\bar{E}_{HO}=0.0723\approx 2/3 {E}_{QS}$, showing that the
systems are indeed in thermal equilibrium.

This is an important result of this paper, that distinguishes between
the coupling of the HO with a single QS (where results are averaged
over many realizations with different initial conditions)
\cite{bonanca1} and the coupling with many ($N \simeq 100$)
simultaneous QS (and no average over realizations). In the former case
a temperature can be defined only by modifying the definition of
entropy \cite{adib,bianucci}. When several QS's are present, the
indirect, and therefore weak, interaction between the members of the
environment plays a key role in the dynamics of the system. It is this
weak interaction that allows for their eventual equilibration.

These numerical results suggest that classically irreversible processes
may indeed occur in nature if the system interacts with a small, but
not too small, chaotic environment. The primary coupling between the
system and the environment leads to dissipation whereas the secondary
interactions between the few, but chaotic, degrees of freedom of the
environment, leads to thermalization.

\section{Linear response}

We can understand the basic mechanisms of dissipation using linear
response theory (\emph{LRT}) \cite{kubo2}. We start by writing an
approximation to the full Hamiltonian of the system as
\begin{displaymath}
H=H_E(Q,P)+H_{I}(Q,P,t)
\end{displaymath}
where $Q$ and $P$ represent the full set of canonical variables of the
environment, $\{x_n,y_n,p_{x_n},p_{y_n}\}$, $n=1,\dots N$. The second term
$H_I(Q,P,t)\equiv A(Q,P)\chi(t)$ is a perturbation that includes the
central system through the time dependent function $\chi(t)$. This
approximation to \eqref{eq1} is valid in the limit of weak coupling, since
the central system is treated as an external source that does not respond
to the environment. The feedback of the environment on the central system
will be considered in a moment.

The dynamics can be described by the Liouville equation
\begin{equation}
\frac{\partial\rho}{\partial t}=iL_{E}\rho(t)+iL_{I}(t)\rho(t).
\label{eqliou}
\end{equation}
where the Liouville operators $L_E$ and $L_I$ are given by
\begin{equation}
iL\rho={\displaystyle \sum_{j}\left[\frac{\partial H}{\partial Q_{j}}
\frac{\partial\rho}{\partial P_{j}}
-\frac{\partial H}{\partial P_{j}}\frac{\partial\rho}{\partial Q_{j}}\right]
=i}\left\{ H,\rho\right\} .
\label{eqliouint}
\end{equation}
with $H$ replaced by $H_E$ and $H_I$ respectively. Eq.\eqref{eqliou}
can also be written in integral form as
\begin{equation}
\rho(t)=e^{i(t-t_{0})L_{0}}\rho(t_{0})+i\int_{t_{0}}^{t}
e^{i(t-s)L_{0}}L_{I}(s)\rho(s)ds.
\label{eqliousol}
\end{equation}

The initial distribution $\rho(t_0)$ is assumed to be invariant under
$H_E$, like, for example, the pseudo-microcanonical or pseudo-canonical
distributions given by Eq.\eqref{eqdist}. The action of $H_{I}(t)$
removes the environment from this initial equilibrium. If the
perturbation is small, Eq.\eqref{eqliousol} can be expanded to first
order in $L_I$ as
\begin{eqnarray}
\rho(t) & = & \rho(t_0)+\int_{t_{0}}^{t}e^{i(t-s)L_{0}}
\left\{ H_{I}(s),\rho(t_0)\right\} ds.
\label{eqlioupert}
\end{eqnarray}
The ensemble average of a general function $B(Q,P)$ can then be calculated
as
\begin{equation}
\langle B(Q,P)\rangle(t)=\langle B(Q(t),P(t))\rangle_0+
\int_{t_{0}}^{t}\phi_{BA}(t-s)\chi(s)dQdPds.
\label{eqB}
\end{equation}
where
\begin{equation}
\phi_{BA}(t)=\langle\left\{ A(Q,P),B(Q(t),P(t))\right\} \rangle_0
\label{eqresp}
\end{equation}
is the \emph{response function} of $B(Q,P)$ when the environment is
under the influence of the perturbation $A(Q,P)$. The subscript $0$
indicates that the averages are computed with the initial invariant
distribution $\rho(Q,P,t_0)$.

We can now compute the response of the central system to the
environment. For the Hamiltonian \eqref{eq1} the HO satisfies the
equation
\begin{equation}
\ddot{q}+\omega_{0}^{2}q=-\frac{\lambda_{N}}{m}\sum_{n=1}^{N}x_{n}
\equiv \frac{\lambda_{N}}{m}X(t)
\label{eq:eqmov}
\end{equation}
and is perturbed by the `external force' $F(t)=\frac{\lambda_{N}}{m}X(t)$.
If the dynamics of the bath coordinates $x_n(t)$ are chaotic we may
replace $X(t)$ by its average $\langle X(t)\rangle$:
\begin{equation}
\ddot{q}+\omega_{0}^{2}q \approx
-\frac{\lambda_{N}}{m}\langle X(t)\rangle.
\label{eq:eqmov medio}
\end{equation}
\begin{figure*}[ht]
\begin{centering}
\includegraphics[scale=0.19]{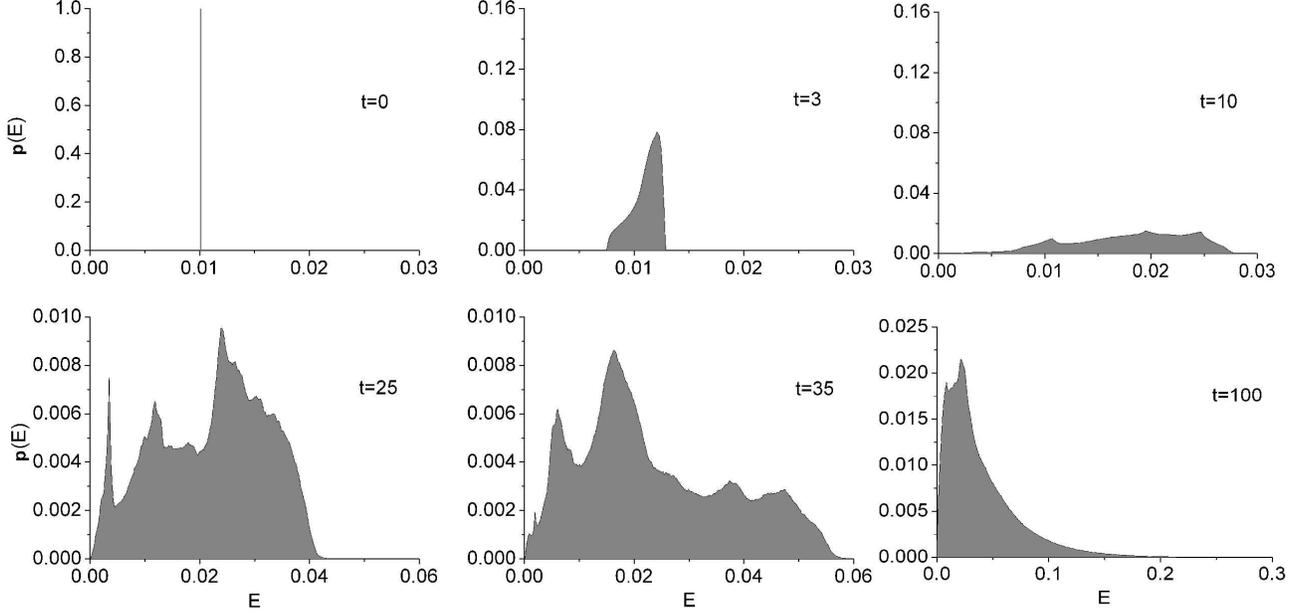}
\par
\end{centering}
\caption{Time evolution of the  environment energy distribution. At t=0
$\rho_{E}$ is given by equation \eqref{eqdist} with $E_{Cn}=0.01$ for all
QS's and $N=100$. Note the different scales in the panels.}
\label{fig5}
\end{figure*}

Using Eq.\eqref{eqB},
\begin{equation}
\langle X(t)\rangle
=\langle\sum_{n}x_{n}(t)\rangle_0-\lambda_{N}\int_{0}^{t}\Phi_{XX}(t-s)q(s)ds.
\label{eq:x medio final}
\end{equation}
where $\Phi_{XX}(t-s) = \left\langle \left\{ X(t),X(s)\right\}
\right\rangle_0$. The first term in the righthand side of Eq.\eqref{eq:x
medio final} is zero due to the parity of $H_E$. Using the
pseudo-microcanonical as the initial invariant distribution it follows
that
\begin{equation}
\Phi_{XX}(t-s)=\sum_{n=1}^{N}\Phi_{xx}^{(n)}(t-s) = N \, \Phi_{xx},
\label{eqresptot}
\end{equation}
where
\begin{eqnarray}
\Phi_{xx}(t-s) & = & \frac{5}{4E_{c}(0)}\left\langle x(t)P_{x}(s)\right
\rangle _0+\nonumber \\
&  & +\frac{t-s}{4E_{c}(0)}\left\langle P_{x}(t)P_{x}(s)\right\rangle _0 ,
\label{eqrespind}
\end{eqnarray}
is the Response Function obtained by Bonan\c{c}a \cite{bonanca1} for
the case of a single QS coupled with the oscillator. Substituting
\eqref{eqresptot} in \eqref{eq:x medio final} and \eqref{eq:eqmov
medio} we obtain
\begin{equation}
\ddot{q}+\omega_{0}^{2}q \approx
\frac{\lambda_{N}^2 N}{m} \int_{0}^{t}\Phi_{xx}(t-s)q(s)ds,
\label{eq:eqmov medio2}
\end{equation}
that shows that it is necessary to re-scale the coupling constant as
$\lambda_{N} = \lambda/\sqrt{N}$, in order for the external driving
$F(t)$ to be independent of the size of the environment in the limit of
large $N$.

The approximation where the $N$ chaotic systems forming the environment
are treated independently correspond to ensemble averages of a single
QS coupled to the central system. As we have seen, this approximation
cannot describe the long time equilibration of the environment, since
this requires some interaction between the chaotic systems. Linear
response theory cannot, therefore describe the long-time behavior of
the system. In particular, it cannot describe the change from an
initial pseudo-microcanonical distribution to the final canonical
distribution.

\begin{figure}[ht]
\begin{centering}
\includegraphics[scale=0.35]{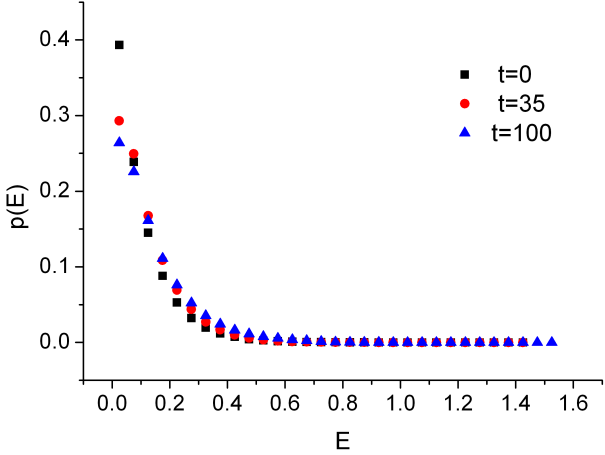}
\par
\end{centering}
\caption{Energy distribution of the environment at different times for
$N=100$. At $t=0$ $p(E)= \exp{(-E/\bar{E}_{QS})}/\bar{E}_{QS}$ with
$\bar{E}_{QS} = 0.01$.}
\label{fig6}
\end{figure}
Fig.\ref{fig5} shows the breakdown of the pseudo-microcanonical
distribution and its consecutive rearrangement to a Boltzmann-like
distribution as a function of time. The transition between the
distributions occurs at about  $1/10$ of the characteristic time for
dissipation, as can be seen from Figs.\ref{fig5} and \ref{fig10}. As
pointed out in the last section, this redistribution of energy is
promoted by the secondary interactions between the QS's.

This rapid equilibration on the environment hints to the possibility of
using LRT to describe the long time behavior of the system by changing
from the initial pseudo-microcanonical to the pseudo-canonical
distribution. If the initial temperature, as given by $\beta_{QS} =
\bar{E}_{QS}$, does not change much during the time evolution, the
environment will not be largely affected by the central system, and the
interaction between the QS's can be ignored. Fig.\ref{fig6} shows an
example with $N=100$ and $\bar{E}_{QS}(0) = 0.01$ where the
distribution of energy of the environment does not change significantly
with time.

Therefore, for $N$ sufficiently large, we can establish a regime where
the central system dissipates energy exponentially, but the
environment's temperature remains nearly constant. This allows us to
use the LRT to describe the system dynamics for much longer times. In
order to do this we need to obtain explicit expressions for
Eqs.\eqref{eqresptot} and \eqref{eqrespind} for the pseudo-canonical
distribution. We write the correlation functions in \eqref{eqrespind}
as
\begin{eqnarray}
\sum_{n}x_{n}(t)P_{x_{n}}(s) & =
& \frac{d}{ds}\sum_{n}x_{n}(t)x_{n}(s)\label{eq:xp}\\
\sum_{n}P_{x_{n}}(t)P_{x_{n}}(s) & =
& \frac{d^{2}}{dtds}\sum_{n}x_{n}(t)x_{n}(s)
\label{eq:pp}
\end{eqnarray}
and
\begin{eqnarray}
\Phi_{XX}(t-s) & = & \frac{5}{4}\frac{d}{ds}\left\langle
\sum_{n}\frac{1}{E_{c_{n}}(0)}x_{n}(t)x_{n}(s)\right\rangle _0+\nonumber \\
 &  & +\frac{(t-s)}{4}\frac{d^{2}}{dtds}\left\langle
 \sum_{n}\frac{1}{E_{c_{n}}(0)}x_{n}(t)x_{n}(s)\right\rangle _0
 \label{eq:funcao resposta}
 \end{eqnarray}

\begin{figure}[h]
\begin{centering}
\includegraphics[scale=0.35]{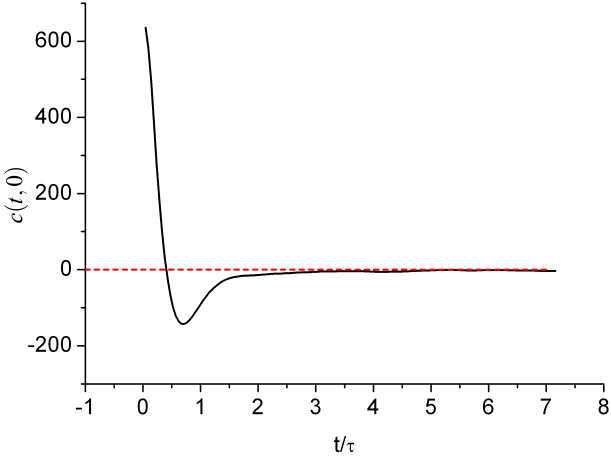}
\par
\end{centering}
\caption{The total correlation function
$C(t,0)=\sum_{n}c_n(t,0)$ with $n=1,\ldots,N=100$
QS's obeying a pseudo-canonical distribution with $\bar{E}_{QS} = 0.01$}
\label{fig7}
\end{figure}

The individual autocorrelation functions $c_{n}(t,s)= \frac{1}{E_{Cn}}
\left\langle x_n(t)x_n(s)\right\rangle _0$ were obtained numerically
and follow the typical behavior expected for chaotic systems
\cite{anishchenko} which can be fitted by $c_n(t,s)=\sigma_n
e^{-\xi_n(t-s)}\cos[\nu_n(t-s)]$. However, the total correlation
function, $C(t,s) = \left\langle\sum_{n}
\frac{1}{E_{c_{n}}(0)}x_{n}(t)x_{n} (s) \right\rangle_0$, smooths out
the oscillatory behavior exhibited by a single QS, as shown in
Fig.\ref{fig7}. It is important to note that the time scale where
$C(t,s)$ is significant is much smaller than the time scale of the
dissipation, as can be seen by comparison with Fig.\ref{fig2}.

\begin{figure}[h]
\begin{centering}
\includegraphics[scale=0.35]{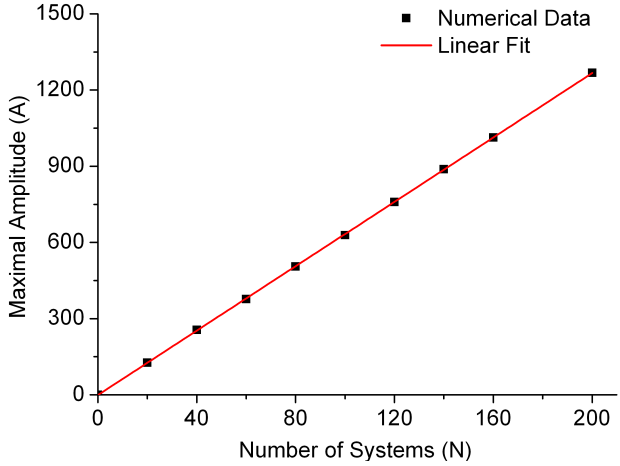}
\par
\end{centering}
\caption{Linear dependence between the maximal amplitude of the total
correlation function $C(t,0)$ and the number of QS's in the environment.
The slope depends on the environment's mean energy.}
\label{fig8}
\end{figure}

Therefore, in the time scale of dissipation, we can make the
approximation
\begin{equation}
\left\langle \sum_{n=1}^{N}\frac{1}{E_{c_{n}}(0)}x_{n}(t)x_{n}(s)
\right\rangle _0 \approx \kappa_{\bar{E}} \mu_{\bar{E}} N \delta(t-s)
\label{eq:correlacao-delta}
\end{equation}
where the parameter $\kappa_{\bar{E}}$ will be adjusted later and $A_N
= \mu_{\bar{E}} N$ is the maximal amplitude of $C_N (t,s)$. It turns
out that $A_N$ increases linearly with $N$, as shown in Fig.\ref{fig8},
and we call $\mu_{\bar{E}}$ the slope of such linear function. The
index ``$\bar{E}$" on $\kappa_{\bar{E}}$ and $\mu_{\bar{E}}$ expresses
the dependence with the mean energy of reservoir, as we will see in a
moment.

Equation \prettyref{eq:correlacao-delta} indicates that the presence of
chaos introduces a fast memory loss in the microscopic dynamics of
reservoir, so that its dynamics can be described by a Markovian
process, unlike the integrable and mixed regimes.

\begin{figure}[h]
\begin{centering}
\includegraphics[scale=0.35]{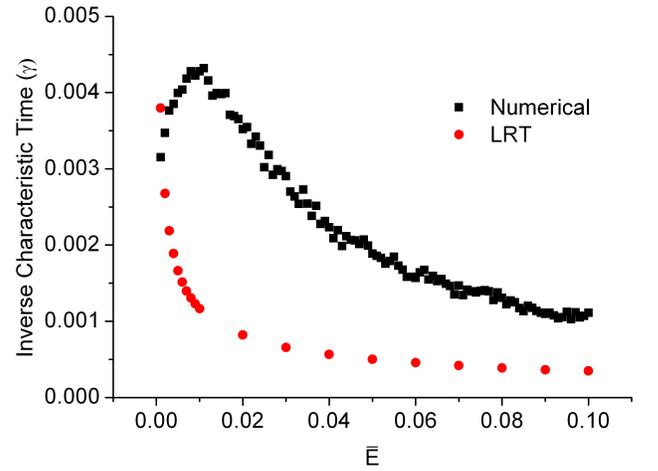}
\par
\end{centering}
\caption{\label{fig9}The inverse characteristic time as a function of the
environment's mean energy for: theoretical model ($\gamma_{\bar{E}}$, red
circles); and numerical data ($\gamma$, black squares). }
\end{figure}

\begin{figure}[h!]
\begin{centering}
\includegraphics[scale=0.35]{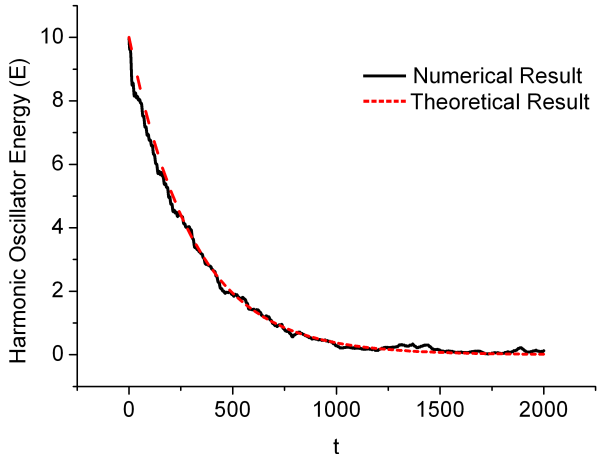}
\par
\end{centering}
\caption{\label{fig10}Comparison between the numerical results
(continuum black line) and theoretical model (dashed red lines), for
$N=100$ and $\kappa_{\bar{E}}=3.68$.}
\end{figure}

Using \prettyref{eq:correlacao-delta} we obtain the response function
\begin{eqnarray}
\phi_{XX}(t-s)&=&\frac{5\kappa_{\bar{E}} \mu_{\bar{E}} N}{4}\frac{d}{ds}
\delta(t-s) \nonumber\\
&&+\frac{\kappa_{\bar{E}} \mu_{\bar{E}} N\left(t-s\right)}{4E_{C_{n}}(0)}
\frac{d^{2}}{dtds} \delta(t-s).
\label{eq:funcao resposta final}
\end{eqnarray}
Substituting $\phi_{XX}(t-s)$ in equations \prettyref{eq:eqmov medio}
and \prettyref{eq:x medio final}, and computing the integrals, we
obtain
\begin{equation}
\ddot{q}+\omega_{o}^{2}q  +\gamma_{\bar{E}}\dot{q}= 0
\label{eq:HO}
\end{equation}
where

\begin{equation}
\gamma_{\bar{E}}=\frac{7\lambda_{N}^{2}\kappa_{\bar{E}} \mu_{\bar{E}} N}{4m}
=\frac{7\lambda^{2} \kappa_{\bar{E}}\mu_{\bar{E}}}{4m}.
 \label{gamma}
\end{equation}
The inverse characteristic time $\gamma_{\bar{E}}$ obtained from LRT
decreases exponentially with $\bar{E}$ as shown in Fig.\ref{fig9}
(circles). The numerically obtained values, on the other hand, displays
a more complex behavior, as shown by the squares in the same figure.
The numerical results can be divided into two regimes: the first, where
$0<\bar{E}<0.01$ and $\gamma(E)$ increases with $E$ and, the second
regime, where $\bar{E}>0.01$ and $\gamma(E)$ decreases with $E$.

The discrepancies between the numerical and LRT results can be
understood if we note that in the first regime, where $\bar{E}$ is
small, the QS's are very slow. Since the value of $\gamma_{\bar{E}}$ is
related to the maximal amplitude of the correlation $C(t,s)$, this
value increases considerably in these regimes because changes in the
state of the quartic oscillators are slow. This, in turn, implies a
super-estimated value for $\gamma_{\bar{E}}$. In this regime, as
$\bar{E}$ increases, the influence of the environment on the HO
increases, facilitating the flow of energy and increasing
$\gamma_{\bar{E}}$. This is a non-trivial effect, since an environment
with lower mean energy should correspond to a cooler thermal bath, with
a greater dissipation rate. This shows that the finite set of QS's
cannot be traded by a thermal bath at very low energies.

After a threshold is passed $\gamma$ recovers the expected decreasing
behavior and the association between the mean energy and temperature is
restored. Even in this case, where the qualitative behavior of
$\gamma_{\bar{E}}$ and $\gamma$ is the same, the numerical value of the
coefficients needs some correction. We define the parameter
$\kappa_{\bar{E}}=\gamma/\gamma_{\bar{E}}$ in order to adjust the
theoretical model with numerical results. When this is done the
numerical and theoretical curves fall on top of each other, as shown in
Fig.\ref{fig10}.

\section{Conclusions}

We constructed a simple model of classical dissipation where a harmonic
oscillator interacts with an environment consisting of $N$ identical
quartic sub-systems (QS) with two degrees of freedom. Interactions
among the sub-systems occur via the harmonic oscillator and are,
therefore, of second order in the coupling constant. In order to
compare results with different number of QS's we re-scaled the coupling
constant according with $\lambda_N = \lambda/\sqrt{N}$.

The main result of this paper is that a small chaotic environment can
indeed behave as an effective infinite reservoir, promoting energy flow
and equilibration with smaller systems in a single realization of the
dynamics. This is to be contrasted with previous results where a single
chaotic system plays the role of environment and thermodynamical
behavior is achieved via ensemble average over many realizations with
random initial conditions \cite{bonanca1}. For the same values of
parameters the results with $N=100$ are roughly equivalent to averaging
over 30,000 initial conditions. Comparing our model environment with
those composed of harmonic oscillators with linear distribution of
frequencies \cite{jane1,jane2}, the number of oscillators needed to
mimic an infinite reservoir is at least an order of magnitude larger
than needed for chaotic QS's.

At long times the environment and the HO equilibrate and their energy
distributions converge to Boltzmann-like curves with the same
temperature. The equilibration of the environment depends on the indirect
interactions among its members that occur via the HO.

We have also developed a Linear Response Theory to describe the system.
We have shown, in particular, that the scaling of the coupling constant
with $N$ is indeed correct and that dissipation constant $\gamma$
depends quadratically on $\lambda$. \\ \\

\centerline{\bf{Acknowledgments}}

\noindent This work was partly supported by Fapesp and CNPq. We thank
M.V.S. Bonan\c{c}a and T.F. Viscondi for many useful discussions.


\end{document}